\documentclass[aps,prd,superscriptaddress,11pt,showpacs,notitlepage]{revtex4}
\pdfoutput=1
	\usepackage{graphicx}
	\usepackage{amsmath,amssymb,mathrsfs}
	\usepackage[colorlinks=true, a4paper=true, pdfstartview=FitV,
linkcolor=blue, citecolor=blue, urlcolor=blue]{hyperref}

\usepackage{epsf}
\usepackage{epsfig}
\usepackage[toc,page]{appendix}
\usepackage{color}

\usepackage{amsmath}

\usepackage{amssymb}

\numberwithin{equation}{section}

\newcommand{\be}{\begin{equation}}
\newcommand{\ee}{\end{equation}}
\newcommand{\bea}{\begin{eqnarray}}
\newcommand{\eea}{\end{eqnarray}}

\newcommand{\bb}{\bibitem}
 
\newcommand{\eqn}{\begin{eqnarray}}
\newcommand{\eqnx}{\end{eqnarray}}
\newcommand{\bA}{{\bf A}}
\newcommand{\bB}{{\bf B}}

\begin{document}
\title{Gravitating gauged BPS baby Skyrmions}

\author{M. Wachla}
\affiliation{Institute of Nuclear Physics,  PAN ul. Radzikowskiego 152, 31-342 Krak\'{o}w, Poland}

\begin{abstract}
We show that the minimally gauged BPS baby Skyrme model remains a BPS theory after coupling with gravity that is the topologically nontrivial configurations called baby Skyrmions carrying magnetic flux are solutions to a zero pressure equation. Following that the proper mass, magnetic flux and the proper geometric volume are linear function of the topological charge, while the total ADM mass and geometric radius get a contribution due to the gravitational interaction which is quadratic in the topological charge. All these quantities are found exactly as target space integrals (averages) of the so-called superpotential.  A complete classification of possible mass-radius curves is provided. 

As an example we consider the model with the pion like mass potential, for which, an approximated but analytical form of the superpotential is provided. 
\end{abstract}

\maketitle 
\section{Introduction}
Magnetized gravitating matter in $(2+1)$ dimensions could be viewed as toy models of magnetic compact stars in $(3+1)$ space-time. If the matter can be related to a low energy regime of QCD, for example by means of an effective action, then we arrive at a model of magnetized neutron stars. 

This goal can be realized within the framework of the $(3+1)$ dimensional Skyrme model \cite{skyrme} which is one of the most acceptable effective model of baryons, atomic nuclei and nuclear matter. In fact, in its so-called BPS limit \cite{BPS-Sk} (also \cite{BPS-Sk-1}) the Skyrme model was shown to be able to support neutron stars with observables (maximal mass, maximal radius, mass-radius curve etc.) in a very good agreement with (still poor) observational data \cite{BPS-star}. Such gravitating solitonic solutions go much further than usual charge one gravitating Skyrmion \cite{skyrme-gr} as the topological charge of the maximal mass solution is of the order $10^{57}$. This result was achieved due to the BPS nature of the model and a large moduli of the static solutions forming a group of the volume preserving diffeomorphisms - both closely related to the most crucial features of nuclear matter: its very small binding energies and liquid like nature. Although in such a BPS limit the usual part of the Skyrme model is neglected, the obtained result should give good approximation to the bulk quantities as the BPS part of the full model provides the leading contribution at high density and pressure, which is the case inside neutron stars \cite{BPS-eos}. Of course, at some point also non-BPS part of the Skyrme model should be taken into account. This is a difficult task as Skyrmions in the full model possess very complicated geometric shape with only discreet symmetries \cite{skyrmions} rendering the problem of computation of self-gravitating multi-Skyrmions too complicated. Then, only mean field approach seems to be applicable \cite{piette}. This concerns the vector meson Skyrme model \cite{vecSK} as well as the weakly bound Skyrme model \cite{weakSK}.

In the next step, one should couple the (BPS) Skyrme model with the Maxwell field which requires an introducing of the usual covariant derivative and inclusion of the Maxwell as well as the  WZW term \cite{witten}. Already, the Maxwell contribution (no gravity) breaks all nice properties of the BPS model rendering the analytical computation impossible. 

However, in $(2+1)$ dimensions the situation is much better. First of all, there is a lower dimensional counterpart of the Skyrme model, known as baby Skyrme model \cite{baby}, which also possesses BPS sectors. Especially, the BPS baby Skyrme model \cite{babyBPS}, \cite{baby-Martin}, \cite{stepien}, \cite{bBPS} is a lower dimensional version of the BPS Skyrme model. This BPS theory can be minimally gauged without destroying its BPS nature \cite{g-bBPS}, \cite{g-stepien}. That is to say, that the gauged (strictly specking magnetic) solitons (baby skyrmions) are solutions of certain Bogomolny equations and therefore saturate a pertinent topological bound. Furthermore, the matter is still of a perfect fluid type. Secondly, it has been very recently shown that also coupling to gravity preserves the BPS property of the BPS baby Skyrme model \cite{bBPS-GR}. This allowed for an analytical computation of gravitating baby Skyrmions in the asymptotically flat space-time. Therefore, a natural question arise what happens if both interactions (Maxwell and gravity) are added to the BPS baby Skyrme model simultaneously.  If the gravitating gauged BPS baby Skyrme model remains a BPS theory we get a unique opportunity to study magnetic planar solitons (a toy model of magnetars) in an analytical way. 

This is the main aim of the present work to study the BPS property of the gauged BPS baby Skyrme model after coupling it to gravity. 

\section{The gravitating gauged BPS baby Skyrme model}
\subsection{Static and axially symmetric field equations }

The gravitating gauged BPS baby Skyrme model is given by the following action where the metric tensor is treated as a dynamical quantity
\be
S_{04}=\int d^3 x |g|^{\frac{1}{2}} \left( -\lambda^2 \pi^2 |g|^{-1} g_{\alpha \beta} \tilde{\mathcal{B}}^\alpha \tilde{\mathcal{B}}^\beta - \mu^2 \mathcal{U} - \frac{1}{4e^2}  F_{\mu \nu}^2  \right)
\ee
Here
\be
\tilde{\mathcal{B}}^\mu = \frac{1}{8\pi} \epsilon^{\mu \nu \rho} \vec{\phi} \cdot \left(  D_\nu \vec{\phi} \times D_\nu \vec{\phi} \right)
\ee
is a gauge invariant version of the topological current 
\be
\mathcal{B}^\mu = \frac{1}{8\pi} \epsilon^{\mu \nu \rho} \vec{\phi} \cdot \left(  \vec{\phi}_\nu \times \vec{\phi}_\rho \right)
\ee
The $U(1)$ gauging is preformed in the usual way i.e., by promoting the global $U(1)$ symmetry of the BPS baby Skyrme model to a local one. This means that we change usual derivatives for the covariant versions \cite{g-baby} (for general baby Skyrmions with magnetic field see \cite{g-yasha}, \cite{g-cs})
\be
D_\mu \vec{\phi}= \vec{\phi}_\mu +A_\mu \vec{n} \times \vec{\phi}
\ee
Furthermore, $\vec{\phi}$ i.e., the baby Skyrme field, is a unit three component iso-vector  $\vec{\phi} \in \mathbb{S}^2$ and the constant vector $\vec{n}=(0,0,1)$. Potential $\mathcal{U}$ is assumed to be a one-vacuum potential which depends only on the third component of the matter field. $F_{\mu \nu}$ is the usual field tensor of the $U(1)$ Maxwell field $A_\mu$ in $(2+1)$ dimensions. 

The corresponding Einstein equations are
\be
G_{\alpha \beta} = \frac{\kappa^2}{2} T_{\alpha \beta}
\ee
where $\kappa^2=16\pi G$ and $G$ is 3-dimensional gravity constant. In the subsequent analysis we assume the axial symmetry for the metric
\be
ds^2=\bA (r) dt^2 - \bB (r) dr^2 -r^2 d\varphi^2
\ee
which gives the standard Einstein tensor $G_{\mu \nu}$
\be
G_{00}=\frac{1}{2r} \frac{\bA \bB'}{\bB^2}, \;\;\; G_{11}= \frac{1}{2r} \frac{\bA'}{\bA}, \;\;\; G_{22} = - \frac{r^2}{4} \left( \frac{\bA'}{ \bA} \frac{\bB'}{\bB^2} + \frac{1}{\bB} \left(  \frac{\bA'^2}{ \bA^2} -  \frac{2\bA''}{ \bA}\right) \right)
\ee
This assumption comes from the observation that the gauged BPS baby Skyrme model has ground state solutions (in each topological sector) in such a axially symmetric form. In fact, these energy minimizers enjoy huge degeneracy which is the group of the area preserving diffeomorphisms. 

Next we observe that the energy-momentum tensor has two contributions
\be
T^{\alpha \beta} = T^{\alpha \beta}_{m} + T^{\alpha \beta}_{em}
\ee
where we have the matter part
\be
T^{\alpha \beta}_m  = 2\lambda^2 \pi^2  |g|^{-1}   \tilde{\mathcal{B}}^\alpha  \tilde{\mathcal{B}}^\beta - \left( \lambda^2 \pi^4 |g|^{-1} g_{\mu \nu} \tilde{\mathcal{B}}^\mu \tilde{\mathcal{B}}^\nu - \mu^2 \mathcal{U} \right) g^{\alpha \beta} 
\ee
and the electromagnetic part
\be
T_{em}^{ \alpha \beta} = \frac{1}{e^2} \left( \frac{1}{4} g^{\alpha \beta} F_{\mu \nu} F^{\mu \nu} - F^{\alpha \nu} F^{\beta}_{\; \;\;  \nu}   \right)
\ee
Let us begin with the baby Skyrme contribution. After coupling to gravity the energy-momentum tensor still possesses the perfect fluid form
\be
T^{\alpha \beta}_m = (p+\rho) u^\alpha u^\beta - p g^{\alpha \beta}
\ee
where the proper energy density and pressure are
\be
\rho = \lambda^2 \pi^2  |g|^{-1} g_{\mu \nu} \tilde{\mathcal{B}}^\mu \tilde{\mathcal{B}}^\nu + \mu^2 \mathcal{U} 
\ee
\be
p = \lambda^2 \pi^2  |g|^{-1} g_{\mu \nu} \tilde{\mathcal{B}}^\mu \tilde{\mathcal{B}}^\nu - \mu^2 \mathcal{U} 
\ee
while four velocity 
\be
u^\alpha = \frac{\tilde{\mathcal{B}}^\alpha}{\sqrt{g_{\mu \nu} \tilde{\mathcal{B}}^\mu \tilde{\mathcal{B}}^\nu}}
\ee
For the static configurations it simplifies to 
\be
T^{00}=\rho g^{00}, \;\;\; T^{ij}=-p g^{ij}
\ee
where we also assume no electric field
\be
A_\mu=(0, A_1 (\vec{x}), A_2(\vec{x}))
\ee

Now, consistently with the assumption on the metric we will restrict ourselves to axially symmetric matter and gauge field. This means that
\be
A_0=A_r=0, \;\;\; A_\phi=n a(r)
\ee
while for the baby Skyrme field expressed by the stereographic projection
\be
\vec{\phi}=\frac{1}{1+|u|^2} \left( u+\bar{u}, -i ( u-\bar{u}),
1-|u|^2 \right)
\ee
we apply the following ansatz 
\be 
u=f(r)e^{in\varphi}, \;\;\; h=1-\frac{1}{1+f^2}
\ee
All this leads to the following expressions for the baby Skyrme energy density and pressure
\be
\rho = \frac{\lambda^2 n^2}{4\bB r^2} (1+a)^2 h_r^2 +\mu^2 \mathcal{U}, \;\;\; p = \frac{\lambda^2 n^2}{4\bB r^2} (1+a)^2 h_r^2 -\mu^2 \mathcal{U} 
\ee
while
\be
\tilde{B}^0= - \frac{n}{2\pi} (1+a)h_r
\ee
Note that our topological current differs by a factor $1/r\sqrt{\bB} $ from the usual topological charge density $q$. This is a consequence of our convention to extract the metric factor from the anti-symmetric tensor. Hence
\be
 n=\int \mbox{vol}_{\mathbb{R}^2} \frac{1}{r\sqrt{\bB}} B^0 = - \int r\sqrt{\bB} dr d\varphi \frac{n}{2\pi r\sqrt{\bB}} h_r 
\ee
while
\be
n=\int dr d\varphi \; B^0
\ee

In addition, the electromagnetic part of the energy-stress tensor reads (diagonal terms)
\be
T_{em}^{00}= \frac{n^2}{2e^2} \frac{1}{\bA \bB r^2} a_r^2
\ee
\be
T_{em}^{rr} =  \frac{n^2}{2e^2} \frac{1}{ \bB^2 r^2} a_r^2
\ee
\be
T_{em}^{\varphi \varphi} =  \frac{n^2}{2e^2} \frac{1}{ \bB r^4} a_r^2
\ee

Now, we can write the Einstein equations (the prime is the derivative w.r.t. $r$) in a compact form
\eqn
\frac{\bB'}{\bB} & = & \kappa^2r \bB \tilde{\rho} \label{eqB} \\
\frac{\bA'}{\bA} & = & \kappa^2 r  \bB \tilde{p}  \label{eqA}\\
(\tilde{p}\bB)' & = & \kappa^2 r  \mu^2 \bB^2 \mathcal{U} \tilde{p}  \label{eqh}
\eqnx
where matter density and pressure, with the gauge component included, read 
\be
\tilde{\rho} =  \frac{n^2}{2e^2r^2\bB}a'^2+ \frac{\lambda^2 n^2}{4r^2\bB } (1+a)^2 h'^2 +\mu^2 \mathcal{U}
\ee
\be
\tilde{p} =  \frac{n^2}{2e^2r^2\bB}a'^2+ \frac{\lambda^2 n^2}{4r^2\bB} (1+a)^2 h'^2 -\mu^2 \mathcal{U}
\ee
This set of equation has to be supplemented by the pertinent Maxwell equations 
\be
\frac{1}{e^2} \partial_\nu \left( \sqrt{g} g^{\mu \alpha} F_{\alpha \beta} g^{\beta \nu} \right) = J^\mu
\ee
where $J^\mu$ is the current due to the covariant derivative in the matter part of the model. It reads
\be
J^\mu=\lambda^2 \pi^2 |g|^{-1/2} g_{\alpha \beta} \frac{\partial}{\partial A_{\mu} } \tilde{B}^\alpha \tilde{B}^\beta
\ee
Hence,
\be
\frac{n}{e^2} \partial_r \left( \sqrt{\frac{\bA}{\bB} } \frac{a_r}{r}\right) = J^\phi
\ee
where
\be
J^\phi = \lambda^2\pi^2 |g|^{-1/2} g_{00} \frac{\partial}{\partial A_{\phi} } \tilde{B}^0 \tilde{B}^0 = \lambda^2  \sqrt{\frac{\bA}{\bB}} \frac{n}{2} (1+a) \frac{h_r^2}{r}
\ee
Together we get
\be
\frac{n}{e^2} \partial_r \left( \sqrt{\frac{\bA}{\bB} } \frac{a_r}{r}\right) =\lambda^2  \sqrt{\frac{\bA}{\bB}} \frac{n}{2} (1+a) \frac{h_r^2}{r} \label{eqa}
\ee

So, finally we are left with a system of four ordinary differential equations (\ref{eqB}), (\ref{eqA}), (\ref{eqh}), (\ref{eqa}) for four unknown functions: to metric functions $\bB, \bA$, the baby Skyrme profile $h$ and the gauge function $a$. We impose the following boundary conditions which guarantee a nontrivial topological charge
\be
h(r=0)=1, \;\;\; h(R)=0, \;\;\; h_r(R)=0
\ee
\be
a(r=0)=0 , \;\;\; a_r(R)=0
\ee
\be
\bB (r=0)=1
\ee
\be
\bA(r=0)=1
\ee
where the conditions for the derivatives of the Skyrme and gauge field come from the vanishing of the pressure at the compacton boundary. Here $R$ is a geometric size of the solitons i.e., a value of the radial distance at which the matter field reaches its vacuum value. It can be finite (for compactons) or infinite (for usual infinitely extended solitons). 

\subsection{The BPS property}
It is easy to notice that there is a formal solution corresponding to zero pressure condition. Indeed, 
\be
\bA=1 \;\;\; \mbox{ and } \;\;\; \tilde{p}=0
\ee
solve two field equations (\ref{eqA}) and (\ref{eqh}). Now, we have to solve the remaining two equations (\ref{eqB}), (\ref{eqa}) and show that $\tilde{p}=0$ condition really lead to solitonic solutions. First of all let us perform a change of the radial variable and introduce
\be
\frac{dz}{dr} = r\sqrt{\bB} \label{z}
\ee
Then we find that the solitonic matter ($\tilde{p}=0$) and gauge equations take the form
\be
 \frac{n^2}{2e^2}a_z^2+ \frac{\lambda^2 n^2}{4} (1+a)^2 h_z^2 -\mu^2 \mathcal{U}=0 \label{z-eqh}
\ee
\be
\frac{n}{e^2}   a_{zz} =\lambda^2   \frac{n}{2} (1+a) h_z^2 \label{z-eqa}
\ee
Now, we want to show that these equations, in the new variable $z$, are equivalent to the Bogomolny equations for the non-gravitating gauge BPS baby Skyrme model. Hence, we introduce a target space function $W=W(\phi_3)=W(h)$ such that the magnetic field 
\be
H=\epsilon^{12}F_{12}=\frac{na_r}{r\sqrt{\bB}} =na_z
\ee
obeys
\be
H\equiv- e^2\lambda^2W(h) 
\ee
Then, the Maxwell equation is equivalent to 
\be
W_h=- \frac{n}{2} (1+a)h_z 
\ee
To summarize, the unknown function $W$ which satisfies the two equations 
\bea
na_z&=&-e^2\lambda^2 W(h) \label{BOG1} \\
\frac{n}{2} (1+a)h_z&=&-W_h(h) \label{BOG2}
\eea 
must obey a constrain following from the zero pressure equation
\be
\frac{e^2\lambda^4}{2} W^2+\lambda^2 W_h^2=\mu^2 \mathcal{U} (h) \label{super}
\ee
All together, (\ref{BOG1}),  (\ref{BOG2})and (\ref{super}), form a set of equations known to be Bogomolny equations for the gauged BPS baby Skyrme model in the flat space-time (see \cite{g-bBPS} with an identification $\lambda^2 \rightarrow \lambda^2/8$ and $W\rightarrow 8W$). In appendix A we show how to derive equations (\ref{BOG1}) and (\ref{BOG2}), by starting from (\ref{Mass}) equation and using FOEL (First-Order Euler-Lagrange) method. Since we consider potentials with vacuum at $h=0$ (an possible other isolated vacua) the super potential equation enforces boundary conditions at $h=0$. Namely,
$$ W(h=0)=0, \;\;\; W_h(h=0)=0$$
The existence of a solution of this equation on the whole segment $h \in [0,1]$ obeying the boundary conditions is a rather nontrivial problem.
Observe that in our construction the superpotential equation (\ref{super}) does not show up from "nothing" as a necessity condition for the saturation of the Bogomolny bound. Here, it is derived as the zero pressure condition which is the very center of any BPS solution. 

Now, we can consider the proper mass of the matter i.e., the energy of the soliton with the gauge field included. 
\bea
M=\int d^2 x |g|^{\frac{1}{2}} \tilde{\rho} &=& \int dr d\varphi r\sqrt{\bB}\left(  \frac{n^2}{2e^2r^2\bB}a_r^2+ \frac{\lambda^2 n^2}{4r^2\bB } (1+a)^2 h_r^2 +\mu^2 \mathcal{U}  \right) \\
&=& \int dz d\varphi \left(  \frac{n^2}{2e^2}a_z^2+ \frac{\lambda^2 n^2}{4} (1+a)^2 h_z^2 +\mu^2 \mathcal{U}  \right) \label{Mass} \\
&=& 2\pi |n| \lambda^2 \left\langle W_h \right\rangle_{\mathbb{S}^2} = 2\pi |n| \lambda^2 |W(h=1)|\label{Mass2}
\eea
The second line shows that the proper mass is just the static energy functional of the gauged BPS baby Skyrme model in the flat space. 
The last equality comes from \cite{g-bBPS} and the fact that our solutions do obey the Bogomolny equations and therefore the pertinent topological inequality is saturated. As a consequence, the proper mass is a linear function of the modulus of the topological charge as expected for a BPS system. Furthermore, the coefficient is uniquely given by the value of the superpotential at the anti-vacuum i.e., at $h=1$, which knowledge does not require to find a particular solution but can be obtained from the superpotential (that is a target space) equation. In other words, the proper mass of the gravitating BPS baby Skyrmion is given by a geometric quantity.  
\\
One can also observe that superpotential equation depends only on {\it one} dimensionless combination of the coupling constants. Indeed, if we define a new superpotential
\be
\omega=\frac{\lambda}{\mu} W
\ee
then (\ref{super}) can be rewritten as
\be 
\omega^2_h+\beta^2\omega^2=\mathcal{U} \label{super1}
\ee
where the new dimensionalless parameter 
\be
\beta^2 = \frac{e^2\lambda^2}{2}
\ee
Thus,
\be
M=2\pi |n| \lambda \mu \left\langle \omega_h \right\rangle_{\mathbb{S}^2} = 2\pi |n| \lambda \mu |\omega(h=1)| \label{M}
\ee
In the limit of vanishing gauge coupling constant $\beta=0$ we get $\omega_h=\sqrt{\mathcal{U}}$, which leads to the expression for the BPS baby Skyrme model \cite{babyBPS}. For arbitrary $\beta$ the superpotential equation is a rather complicated nonlinear differential equation. However, as we show it in the next section it can be solved approximately with an arbitrary accuracy.

Next, we can use \cite{g-bBPS} and find the total magnetic flux carrying by the baby Skyrmion. Namely,
\be
\Phi=\int d^2 x |g|^{\frac{1}{2}} H = \int dr d\varphi r\sqrt{\bB} \frac{na_r}{r\sqrt{\bB}}  = 2\pi n \int dz a_z =2\pi n a(z_0) \equiv 2\pi n a_\infty
\ee
where $z_0$ is the geometric size of the soliton (in $z$ variable) which is finite for compactons and infinite for usual infinitely extended solitons. Hence, from the Bogomolny equations one can find \cite{g-bBPS}
\be
a_\infty = -1 + \exp \left(-\frac{F(1)}{4} \beta^2\right)
\ee
where 
\be
F(h)=4 \int_0^h \frac{W(h')}{W_{h'}(h')} dh' =4 \int_0^h \frac{\omega(h')}{\omega_{h'}(h')} dh' \label{F}
\ee
is a function of the target space variable again uniquely defined for a model (potential). Hence, again the total flux can be found without solving Bogomolny equations but only by finding the superpotential.  

Analogously, the geometric volume of the solitons reads
\be
V=\int d^2 x |g|^{\frac{1}{2}} = 2\pi z_0 = \pi \frac{\lambda}{\mu} |n|  \exp \left(-\frac{F(1)}{4} \beta^2\right) \int_0^1 \frac{\exp \left(\frac{F(h)}{4} \beta^2 \right)}{\omega_h} dh 
\ee
Of course, $z_0=V/2\pi$ which we will use later on. 

\vspace*{0.2cm}

The remaining piece is the equation for the metric function $\bB$. It can be formally solved (in the $z$ radial variable)
\be
\bB^{-1/2}(z)=1-\frac{\kappa^2}{2} \int_0^z \tilde{\rho}(z')dz'
\ee
Obviously, as the metric function has to be a regular function we get a constrain
\be
1-\frac{\kappa^2}{4\pi} \int_0^z 2\pi \tilde{\rho}(z')dz' >0 \;\;\; \Rightarrow \;\;\; \frac{\kappa^2 M}{4\pi} <1
\ee
As a result, the magnetic gravitating BPS baby Skyrmions exist until a maximal topological charge $n_{max}$ 
\be
n_{max}=\left\lfloor \frac{2}{\lambda \mu  \kappa^2 |\omega(1)|} \right\rfloor \label{maxn}
\ee
This leads to a maximal proper mass, maximal magnetic flux and maximal proper volume of our gravitating solitons. Specifically,
\be
M^{max}= \frac{4\pi}{\kappa^2}
\ee

The proper mass (non-gravitating mass), magnetic flux as well as the proper geometric volume are quantities which are linear in the topological charge. In fact, after the coordinate change they where obtained simply from the non-gravitating gauged BPS baby Skyrme model. However, there are two "observables" which take into account the gravity interaction in a more non-trivial way. They are the total (asymptotic) mass and the radius. As we will see both can be also obtained without knowledge of local form of the solutions i.e., by proper target space integrals.

The total ADM mass reads
\be
M_{ADM}=2\pi \int_0^{r_0} rdr  \tilde{ \rho} (r)  = 2\pi \int_0^{z_0} \frac{dz}{\sqrt{\bB (z)}}  \tilde{ \rho}(z) =2\pi \int_0^{z_0} dz \tilde{ \rho}(z) \left( 1-\frac{\kappa^2}{2} \int_0^z \tilde{\rho}(z')dz' \right)
\ee
Hence,
\be
M_{ADM}=2\pi \int_0^{z_0} dz \tilde{ \rho}(z)-2\pi\frac{\kappa^2}{2}  \int_0^{z_0} dz \tilde{ \rho}(z) \left( \int_0^z \tilde{\rho}(z')dz' \right)
\ee
However, the double integral can be written as
\be
 \int_0^{z_0} dz \tilde{ \rho}(z) \left( \int_0^z \tilde{\rho}(z')dz' \right)= \frac{1}{2}  \left(  \int_0^{z_0} dz \tilde{ \rho}(z) \right)^2
\ee
which gives
\be
M_{ADM}=M- \frac{\kappa^2}{8\pi}M^2 = M\left(1- \frac{\kappa^2}{8\pi}M \right) \label{M-ADM-M}
\ee
where $M$ is the proper mass. Inserting (\ref{M}) we find an exact formula
\be
M_{tot}= 2\pi |n| \lambda \mu |\omega(h=1)| \left(1-  \frac{\kappa^2\lambda \mu}{4}  |n|  |\omega(h=1)| \right) \label{Mtot}
\ee
The total mass gets a correction due to the gravitational interaction which is quadratical with the topological charge. We remark that $M_{ADM}$ grows with $n$ until $n=n^{max}$ where $dM_{ADM}/dn$ vanishes. In other words, the total mass instability occurs exactly at the maximal mass point. Hence, the maximal total mass is  
\be
M^{max}_{ADM}= M_{ADM}(n^{max})=\frac{M}{2}=\frac{2\pi}{\kappa^2} \label{M-ADM-max}
\ee
which is half of the non-gravitating mass. 

It is interesting to notice relations between the ADM and proper mass (\ref{M-ADM-M}), (\ref{M-ADM-max}) are identical as in the non-gauged case. The unique place where the gauge interaction modifies formulas is the value of the superpotential at the anti vacuum $h=1$, which obviously changes if the gauge coupling constant changes. 

\vspace*{0.cm}

Finally, the radius can be computed from
\be
\frac{R^2}{2}=\int_0^R rdr = \int_0^{z_0} \frac{dz}{\sqrt{\bB (z)}} =\int_0^{z_0} dz \left( 1-\frac{\kappa^2}{2} \int_0^z \tilde{\rho}(z')dz' \right)
\ee
Thus,
\be
\frac{R^2}{2}=\frac{V}{2\pi} -  \frac{\kappa^2}{2}\int_0^{z_0} dz \left( \int_0^z \tilde{\rho}(z')dz' \right)
\ee
In order to compute the double integral part we have to turn to the topological bound. First of all let us underline again that the static proper mass (energy) in the new radial coordinate $z$ is identical with the static energy functional of the gauge BPS baby Skyrme model in the ${\it flat}$ space. Then, following the standard derivation of the Bogomolny bound for the gauge BPS baby Skyrme model and using our axial static ansatz we find that (for simplicity we consider positive topological charge and chose the sign of $W$ such that the integral is positive) 
\be
 \int_0^{z} \tilde{\rho}(z') dz' = n \lambda^2 ( W(1)-W(h) - a(z)W(h)) \label{BPSbound}
\ee
where $h$ is understood as a function of $z$.  Derivation of equation (\ref{BPSbound}) is presented in Appendix B.
If we integrate over the full domain of the solution then $z=z_0$ which corresponds to $h=0$. But then $W(0)=0$ and we arrive at the usual total energy expression (divided by $2\pi$).\\
In the next step we use that 
\be
a(z)= -1 + \exp \left(\frac{F(h(z))-F(1)}{4} \beta^2\right)
\ee
Moreover, any integral over the variable $z$ can be change into a target space expression by 
\be
dz=-\frac{n}{2}\frac{1+a}{W_h}dh =-\frac{n}{2} \frac{\exp \left(\frac{F(h)-F(1)}{4} \beta^2\right)}{W_h} dh
\ee
Finally putting all together we find
\be
\int_0^{z_0} dz \left( \int_0^z \tilde{\rho}(z')dz' \right)= \frac{n^2\lambda^2}{2} \mathcal{A}(\beta)  
\ee
where
\be
 \mathcal{A}(\beta) =  \int_0^1 \frac{\exp \left(\frac{F(h)-F(1)}{4} \beta^2\right)}{\omega_h}  \left[  \omega(1)- \exp \left(\frac{F(h)-F(1)}{4} \beta^2\right)\omega(h)  \right]dh
\ee
depends only on the superpotential i.e., on a particular form of the potential. Note that the double integral, and therefore the gravity modification (shrinking) of the radius, is a quadratic function of the topological charge - exactly as in the non-gauge case \cite{bBPS-GR}. 

Now, we can study the mass-radius relation. This can be performed by introducing a new variable  $x=|n|/n_{max} \in [0,1]$. Then we find such a relation in a parametric way 
\be
\left\{
\begin{array}{l}
\label{Parametr}
\cfrac{\kappa^2 M_{ADM}}{2\pi } = x  \left( 2 -x \right)\\
\\
\cfrac{\kappa^2 \mu^2 R^2}{2}=  \cfrac{\mathcal{A}(\beta)}{|\omega(1)|^2}\left( \cfrac{\mathcal{C}(\beta)|\omega(1)|}{\mathcal{A}(\beta)}-x  \right)
\end{array}
\right.
\ee
where
\be
\mathcal{C}(\beta)= \exp \left(-\frac{F(1)}{4} \beta^2\right) \int_0^1 \frac{\exp \left(\frac{F(h)}{4} \beta^2 \right)}{\omega_h} dh
\ee
Similarly to work \cite{bBPS-GR} we define new parameter $\Omega(\beta)$
\be
\Omega(\beta)=\cfrac{\mathcal{C}(\beta)|\omega(1)|}{\mathcal{A}(\beta)}
\ee 
Qualitatively, we obtain the same family of mass-radius curves as in the non-gauge case ($\beta=0$) \cite{bBPS-GR} governed by the value of $\Omega$. For $\Omega=2$ $M_{ADM}$ is a linear function of $R^2$. For $\Omega <2$ the $M_{ADM}-R$ curve turns left at some value of the topological charge (or $x$) which means that the maximal radius does not coincide with the maximal mass. This is the case for $\Omega >2$, where the curve bends right. 

Of course, one obvious question is whether the value of $\Omega$ can cross 2 while $\beta$ is changed. This would lead to a drastic change of the qualitative behaviour of the mass-radius curve. We will investigate this issue taking the old baby potential. 
\section{Example - The pionic mass potential}
\subsection{Superpotential}
As an example we will consider the most popular old baby potential which is a lower dimensional counterpart of the pionic mass potential for the Skyrme model
\be
\mathcal{U}_\pi = \frac{h}{4}
\ee
We have to begin our analysis with the superpotential equation (\ref{super1}) which knowledge is essential for computation of all  quantities
\be
\omega^2_h+\beta^2\omega^2=\frac{h}{4}, \;\;\;\omega(0)=0
\ee
\begin{figure}
\hspace*{-1.0cm}
\includegraphics[height=5.cm]{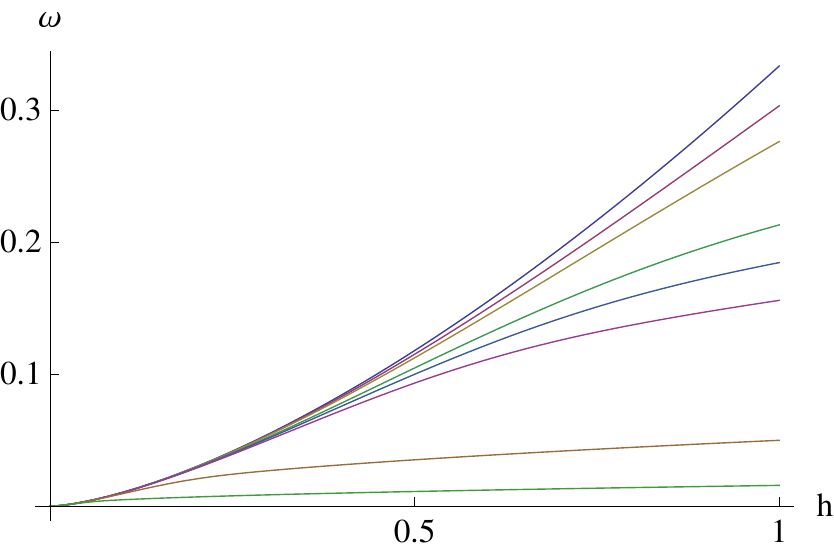}
\includegraphics[height=5.cm]{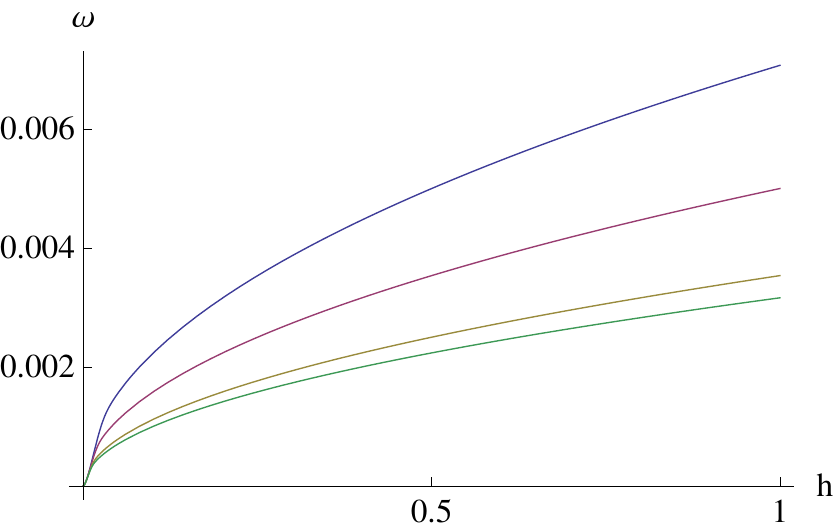}
\caption{Superpotential $\omega$ for the old baby potential. {\it Left panel}: $\beta^2=0,1,2,5,7,10,10^2,10^3$; {\it Right panel}: $\beta^2=5\cdot 10^3, 10^4, 2\cdot 10^4,  2.5\cdot 10^4 $. Increasing $\beta$ corresponds to a more suppressed curve.}
\label{omega}
\end{figure}
Some numerical solutions to this equation in the unit segment $[0,1]$ are presented in Fig. \ref{omega}. There are two limiting cases for which we can solve the equation exactly. 

First, for $\beta=0$ we arrive at the non-gauge case and
\be
\omega_{\beta=0}=\frac{1}{3}h^{3/2}
\ee
For finite but small $\beta$ we may apply the perturbative expansion and find
\be
\omega_{small}=h^{3/2}\left( \frac{1}{3} -\frac{2}{63} (\beta h)^2+\frac{10}{6237} (\beta h)^4-\frac{92}{5893965} (\beta h)^6+o(\beta^6h^6)  \right) 
\ee
which agrees extremely well with numerics for $\beta^2 <8$ on the whole unit segment. For higher value of the parameter the approximated solution begin to disagree in the vicinity of  $h=1$. Once we increase $\beta$ such a disagreement is more and more pronounced and occurs for smaller $h$. This forces us to analyse an expansion at the $h=1$ end.

\begin{figure}
\hspace*{-1.0cm}
\includegraphics[height=5.cm]{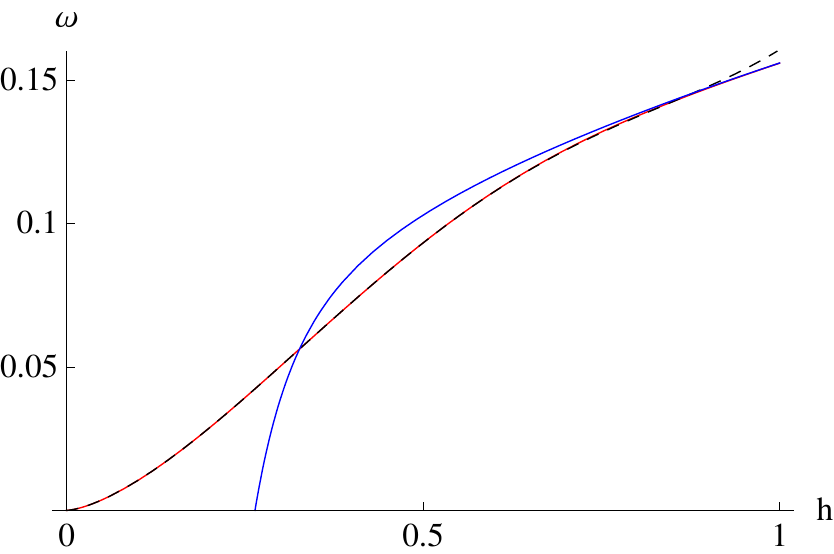}
\caption{The superpotential $\omega$ for the pionic potential and $\beta^2=10$ (red line) with $\omega_{small}$ (dotted) and $\omega_{large}$ (blue) approximate functions.}
\label{approx-beta10}
\end{figure}

Observe first that for very large value of the parameter $\beta \rightarrow \infty$, the superpotential equation  gives  $\omega=\frac{1}{2\beta} h^{1/2}.$ This provides as approximation close to $h=1$. For finite but large $\beta$ we find the following approximated solution
\be
\omega_{large} =  h^{3/2} \left( \frac{1}{2} (\beta h)^{-1} -\frac{1}{16 } (\beta h)^{-3}-\frac{13}{256} (\beta h)^{-5}-\frac{213}{2048} (\beta h)^{-7}  +o(\beta^{-7}h^{-7}) \right)
\ee
Of course, it cannot serve as an approximated a solution on the full segment as its derivative is divergent at the origin. However, at the vicinity of $h=0$ the solution can be always approximate by the small $\beta$ solution. Due to that large $\beta$ approximated solution is 
\be
\omega_{approx}=\left\{
\begin{array}{ll}
\omega_{small} & h \in [0, h_0] \\
\omega_{large} & h \in [h_0,1]
\end{array}
\right.
\ee
where the gluing point is defined as 
\be
\omega_{small} (h_0)=\omega_{large}(h_0)
\ee
which, for the upper assumed order of the expansion, is 
\be
h_0=2.7821 \frac{1}{\beta}
\ee
For too small $\beta$ the gluing point $h_0$ is not in the unit segment and in a consequence the approximated solution is given simply by the small $\beta$ expansion. 
This solution reproduces the true numerical solution with a great accuracy for all $\beta$. In Fig. \ref{approx-beta10} we show the numerical superpotential $\omega$ for $\beta^2=10$ (red curve) together with $\omega_{small}$ (dotted curve) and $\omega_{large}$ (blue curve). The correct approximation $\omega_{large}$ is provided by a composition of the dotted and blue curves glued at the second crossing point $h_0(\beta^2=10)=0.8798$.
\subsection{Masses, magnetic flux, proper geometric volume and radius}
\begin{figure}
\hspace*{-1.0cm}
\includegraphics[height=5.cm]{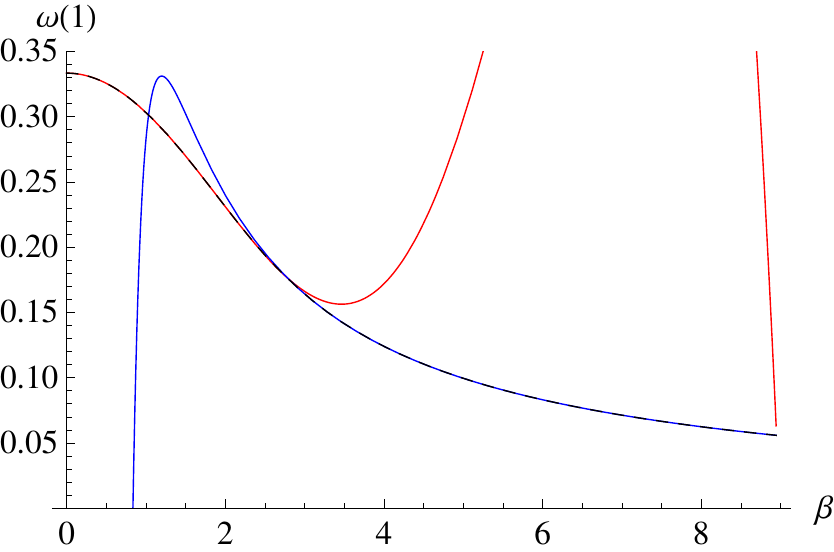}
\caption{The value of the superpotential at $h=1$ i.e., $\omega(1)=M/(2\pi |n| \lambda \mu)$ for the pionic potential as a function of $\beta$. Dashed line - approximated expression; red line - $\omega_{small}(1)$; blue line - $\omega_{large}(1)$.}
\label{omega1}
\end{figure}
\begin{figure}
\hspace*{-1.0cm}
\includegraphics[height=5.cm]{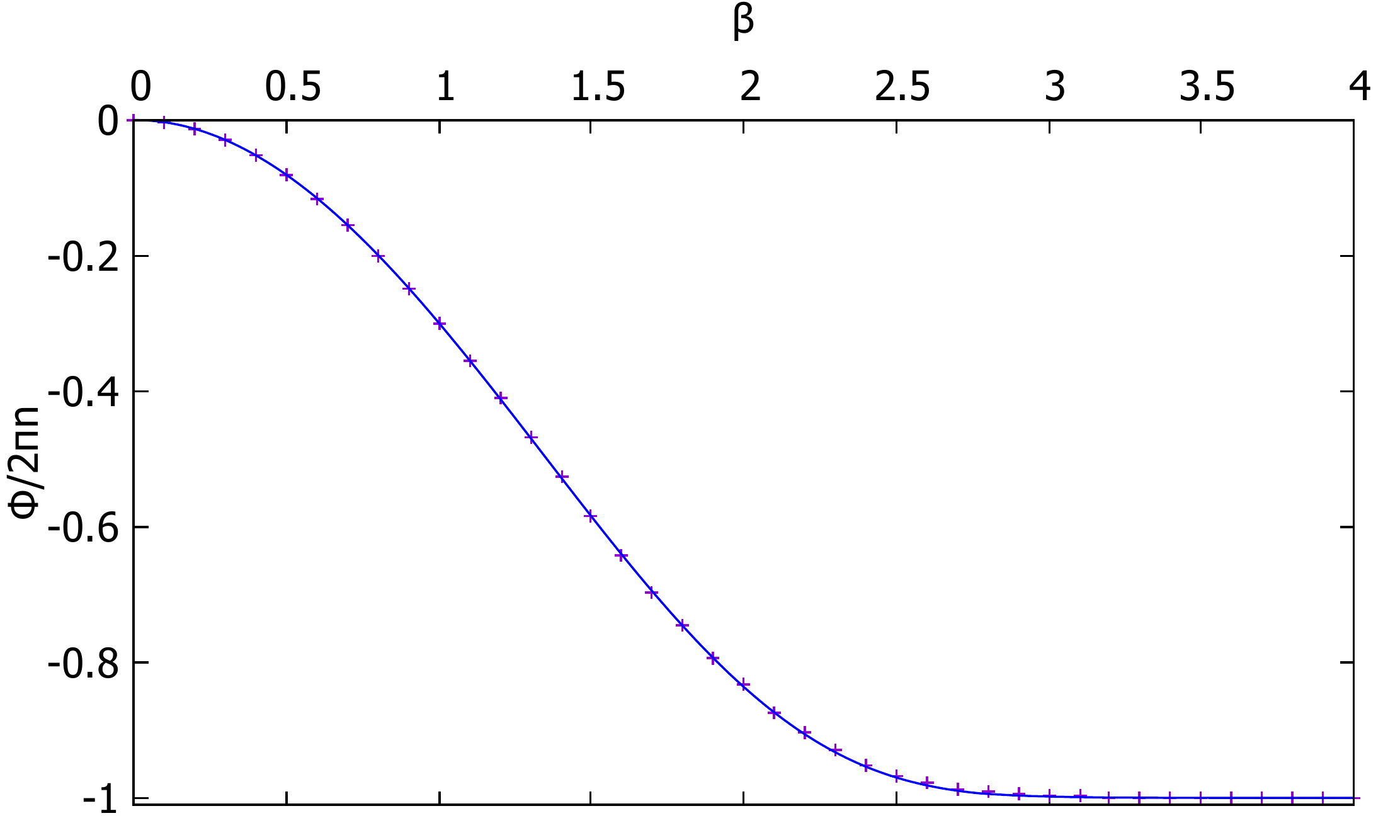}
\caption{The magnetic flux (left, blue line - approximated flux, violet points - numerical flux) for the old baby potential as a function of $\beta$.}
\label{flux}
\end{figure}
\begin{figure}
\hspace*{-1.0cm}
\includegraphics[height=5.cm]{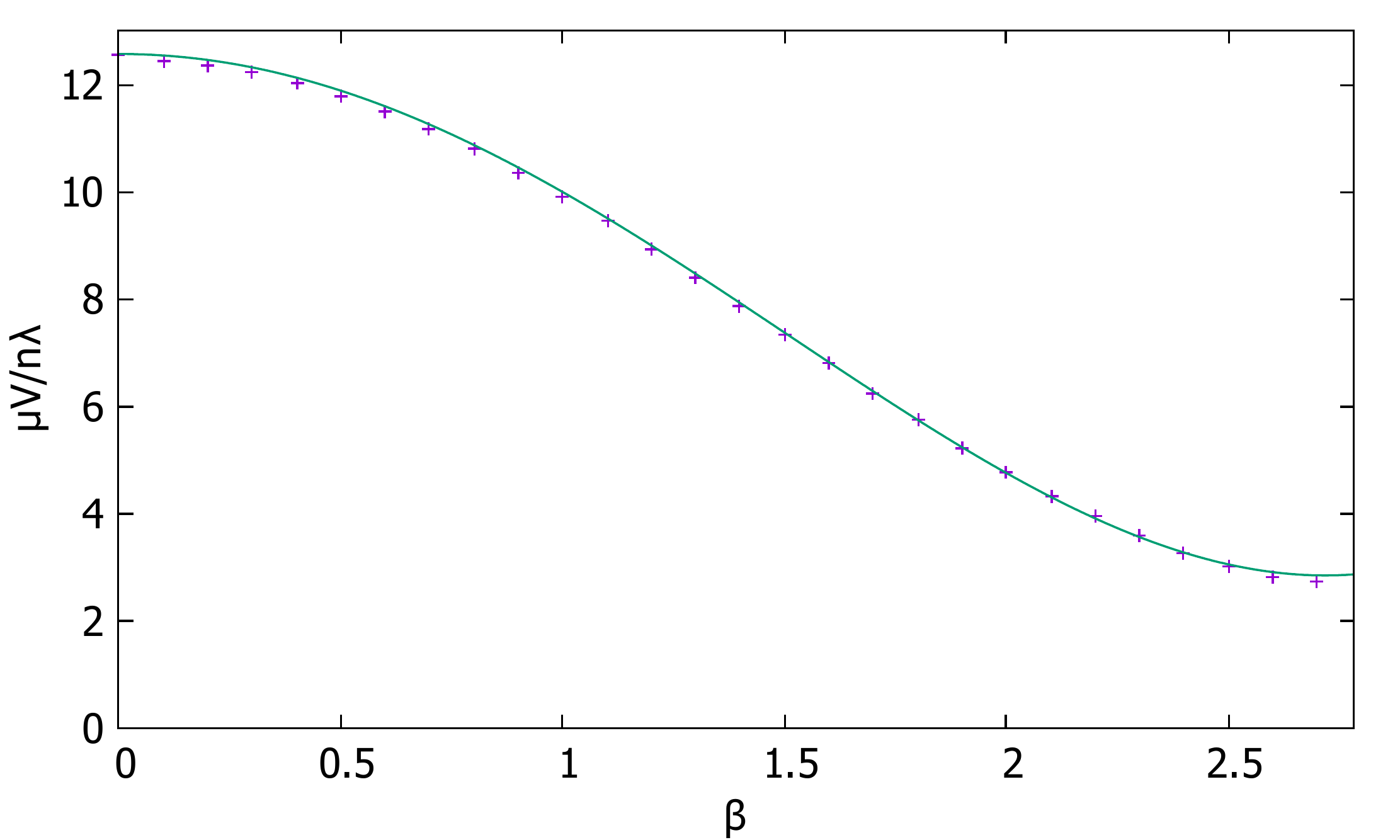}
\caption{The volume for the old baby potential as a function of $\beta$ (violet points - numerical volume, green line - approximated formula).}
\label{V-old}
\end{figure}

Using the value of the approximated superpotential at $h=1$ we can find the following approximated but analytical formula for the proper mass 
\be
M=2\pi |n| \lambda \mu \cdot \left\{ 
\begin{array}{ll}
\frac{1}{3}-\frac{2}{63}\beta^2 +\frac{10}{6237}\beta^4-\frac{92}{5893965} \beta^6 +o(\beta^6) & \beta \le 2.7821 \\
& \\
 \frac{1}{2} \beta^{-1} -\frac{1}{16} \beta^{-3}-\frac{13}{256} \beta^{-5}-\frac{213}{2048} \beta^{-7}  +o(\beta^{-7}) & \beta \ge 2.7821 
 \end{array}
\right. \label{M-old}
\ee
In Fig. \ref{omega1} we plot approximated  $\omega_{approx}(1)$ as a function of the dimensionless parameter $\beta$. The gluing point is $\beta=2.7821$. The true numerical value is undistinguishable from the approximated (dashes) curve. Therefore, the approximated formula for the proper mass agree with the true numerical curve with very good accuracy. 
This formula is also sufficient to get an approximated but analytical expression for the ADM total mass. We just need (\ref{Mtot}). 

In the next step we find an analytical, approximated expression for the magnetic flux. This requires knowledge of $F(h)$ function (\ref{F})
\bea
\frac{1}{4}F(h)&=&\left(\frac{1}{3}h^2+\frac{4}{189}\beta^2 h^4 +\frac{32}{18711}\beta^4h^6+\frac{32}{280 665}\beta^6h^8 \right)\Theta[h_0-h]  \\
&+&\left(-1.76336 \beta^{-2} + \frac{61}{64} \beta^{-6}h^{-4}+\frac{5}{8}\beta^{-4}h^{-2} +h^2-\beta^{-2} \ln h  -1.0232\beta^{-2}\ln \beta  \right) \Theta[h-h_0] \nonumber
\eea
Hence, the flux
\be
\frac{\Phi}{2\pi n} = -1+\exp \left(-\beta^2 \cfrac{F(1)}{4}\right)
\ee
where
\bea
\frac{1}{4}F(1)&=&\left(\frac{1}{3}+\frac{4}{189}\beta^2  +\frac{32}{18711}\beta^4+\frac{32}{280 665}\beta^6 \right)\Theta[2.7821-\beta]  \\
&+&\left(1-1.7634 \beta^{-2} +\frac{5}{8}\beta^{-4} +\frac{61}{64} \beta^{-6}-1.0232\beta^{-2} \ln \beta  \right) \Theta[\beta-2.7821 ] \nonumber
\eea
It is worth to notice that for $\beta=2.7821$ (i.e., when $\omega_{large}$ must be taken into account) the flux is $\Phi/(2\pi n)=-0.9936$, which is very close to its asymptotic value $-1$. The approximated expression for the magnetic flux is plotted in Fig. \ref{flux}.

Although the magnetic flux is practically quantized for $\beta>2.7821$ the proper geometric volume of compactons is still not too small. Specifically, it drops approximately 5 time from the non-gauge case.
An approximated formula for $\beta < 2.7821$ is presented below 
\be
V=4\frac{\lambda}{\mu} |n| \pi \left(1-\frac{2}{9} \beta^2 +\frac{10}{567}\beta^4 -\frac{92}{392931} \beta^6\right)
\ee
We plot it in Fig. \ref{V-old}. 

We conclude that our approximation agrees very well with the numerical results. Of course, taking more terms in the small and large $\beta$ expansion of the superpotential we can approach an arbitrary accuracy, solving the model completely.

\subsection{Mass-radius curve}
\begin{figure}
\hspace*{-1.0cm}
\includegraphics[height=5.cm]{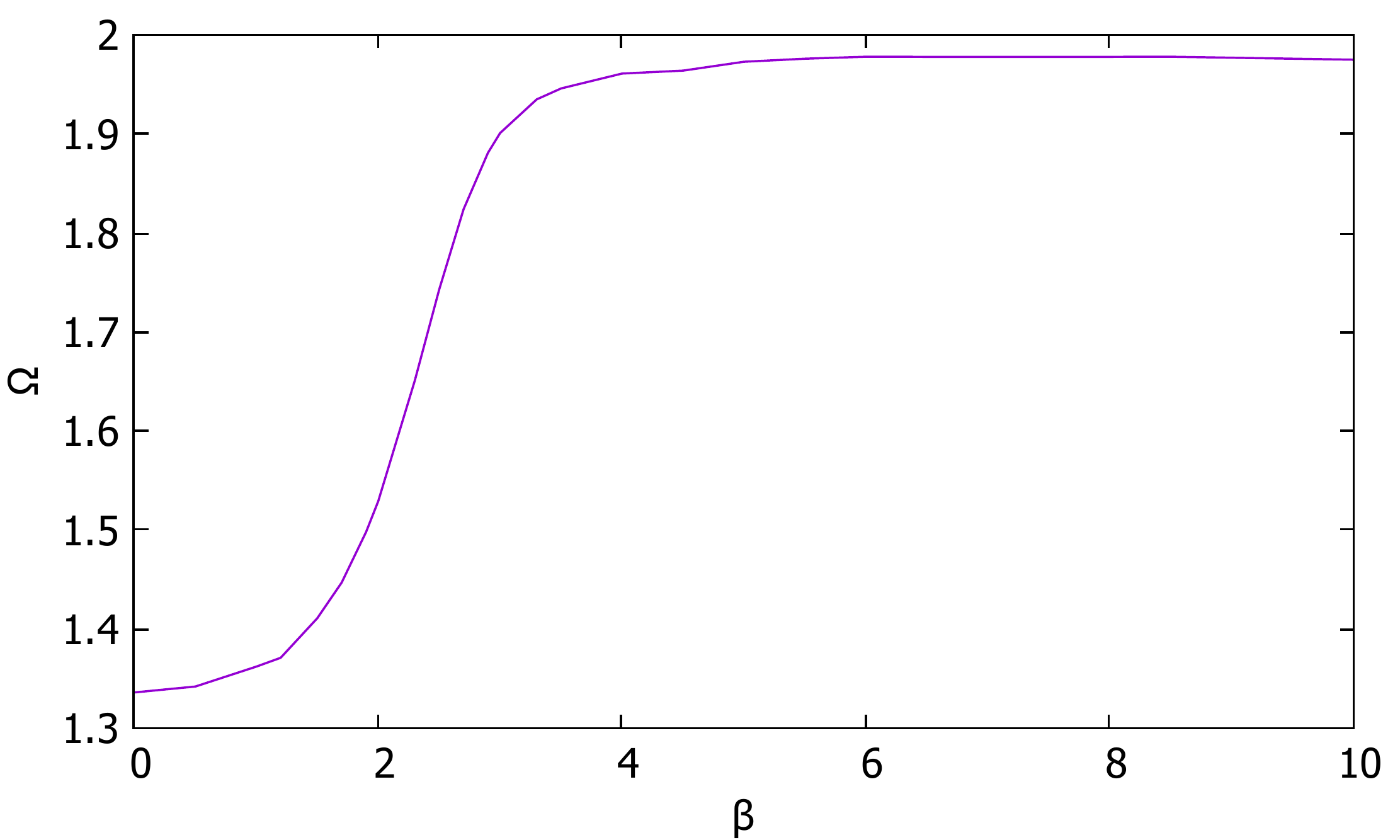}
\caption{$\Omega$ as a function of the coupling constant $\beta$ for the old baby potential.}
\label{OmegaBeta}
\end{figure}

As we know the shape of the mass-radius curve strongly depends on the value of $\Omega$. For the non-gauged model it reads $\Omega(\beta=0)=4/3$. This means that the mass-radius curve bends at some point (the maximal radius point) towards left. In fact, it has been recently observed that in the BPS baby Skyrme model $\Omega < 2$ is a rather preferred value for many one vacuum potentials \cite{bBPS-GR}.  It is a matter of fact that for the old baby potential $\Omega (\beta)$ is a growing function of the coupling $\beta$ - see Fig. \ref{OmegaBeta}. It goes from $4/3$ and asymptotically reaches $\Omega=2$. This can be proven using the approximated superpotential. In fact, if $\beta \rightarrow \infty$, it is enough to take $\omega=\omega_{large}=\sqrt{h}/(2\beta)$. Then, all possible corrections (from $\omega_{small}$) contribute to $h\rightarrow 0$ end, which, due to the regularity of the integral does not have any importance for the value of $\Omega(\beta=\infty)$. 

\begin{figure}
\hspace*{-1.0cm}
\includegraphics[height=5.cm]{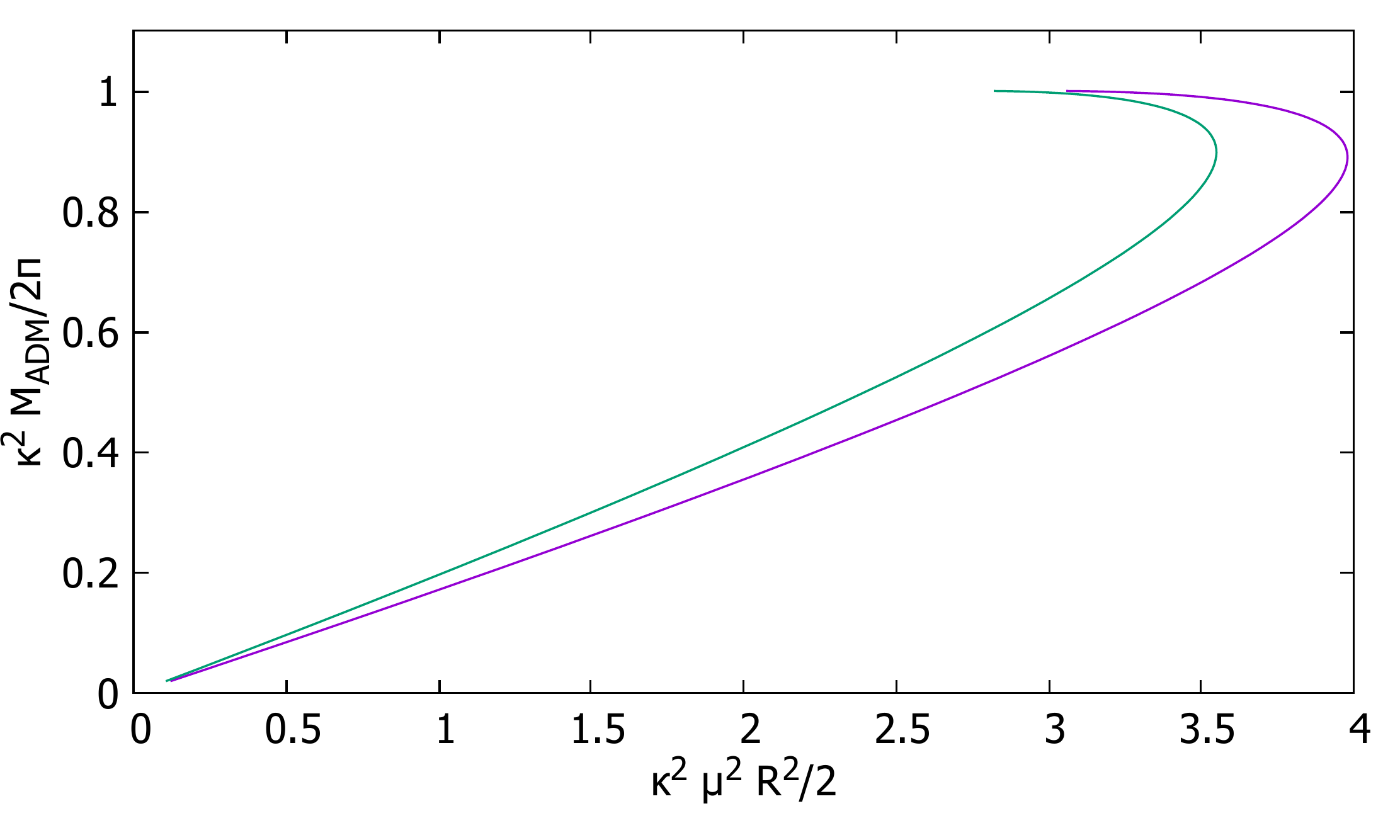}
\includegraphics[height=5.cm]{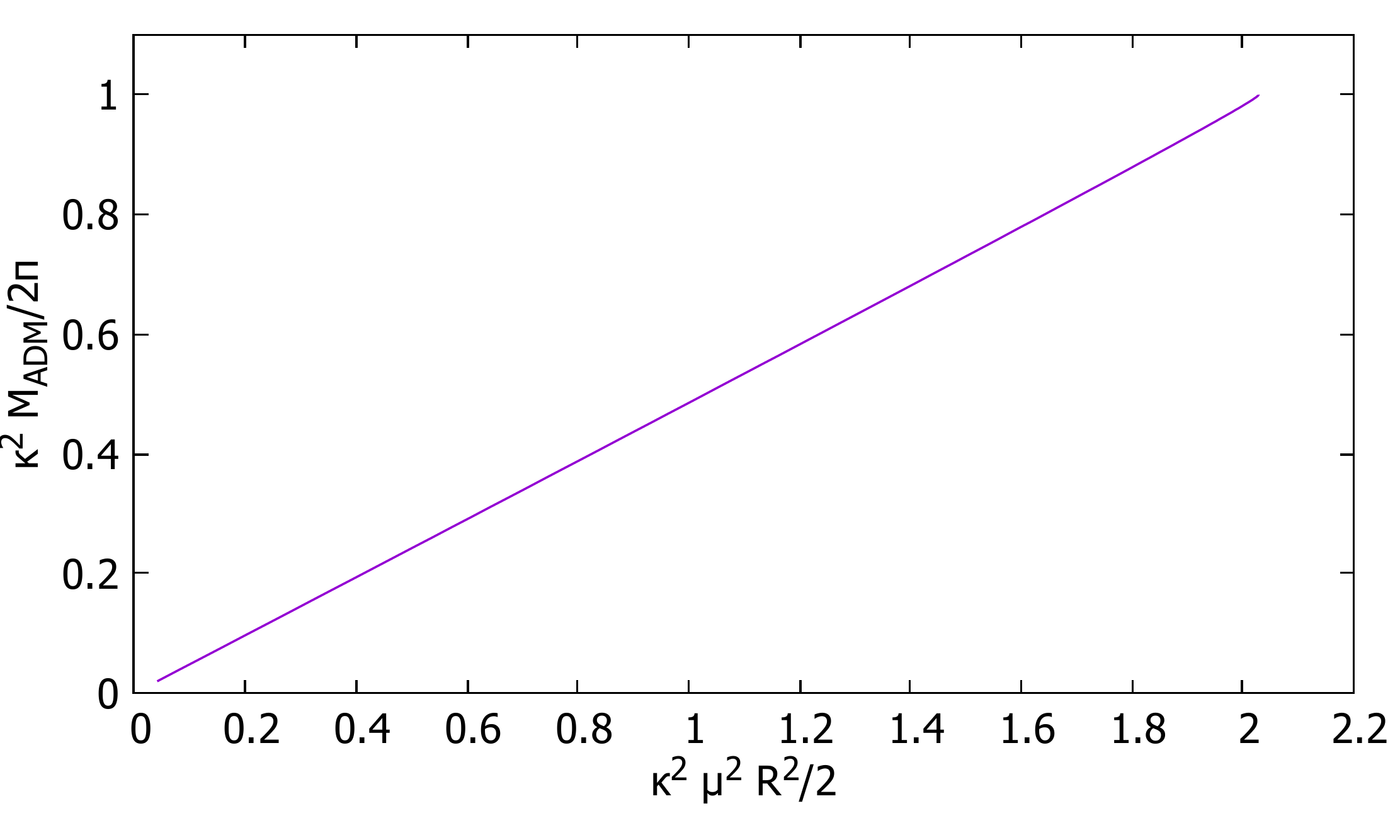}
\caption{The mass-radius square curve. Left: $\beta=0$ (violet) and $\beta=1$ (green). Right: $\beta=4$.}
\label{MR}
\end{figure}

A physical explanation of this asymptotical behaviour of $\Omega$ is quite obvious. As the coupling constant grows solitons become more and more squeezed which physically means that the matter is more and more stiff with the energy density given by almost a step function. However, it is known that for the maximally stiff mater i.e., the BPS baby Skyrme model with the Heaviside step potential, $\Omega=2$ and we arrive at the linear dependence between mass and radius squared. In Fig. \ref{MR} we plot the mass-radius curve for $\beta=0$ (non-gauge case with $\Omega=4/3$), $\beta=1$ ($\Omega=1.36$) and $\beta=4$ ($\Omega=1.96$).  

\vspace*{0.2cm}

Let us notice that the magnetic interaction lowers the proper mass. A related observation is that increasing of $e$ increases the value of the maximal topological charge carried by the gravitating soliton. It is because  $\omega(1)$ gets smaller in  (\ref{maxn}).

\section{Summary}
In this paper we investigated BPS baby Skyrme model in (2+1) dimensions coupled simultaneously to the Maxwell field and gravity. Such a theory, as we argued in the Introduction, might be used as a toy model of magnetised neutron stars. 

The first main result is that such a self gravitating theory of magnetised nonlinear matter with a nontrivial topology remains a BPS theory, that is it supports solitonic solutions, being magnetised gravitating baby Skyrmions, as solutions of a zero pressure equation i.e., it admits a reduction to a Bogomolny equations where the proper (nongravitational) mass is a {\it linear} function of the topological charge. Corresponding topological lower bound on the proper mass integral is saturated. In a consequence we derive a theory of magnetised and gravitating perfect fluid solitons in (2+1) dimension which is completely solvable, in the sense that all observables are given as some functions of topological charge with constants being target space integrals depending on the coupling constant $\beta=e\lambda/\sqrt{2}$ and a particular model (particular potential). Hence, all observables are computable without any knowledge of the local form of solutions. 

Specifically, the proper mass and proper geometric volume are linear function of the topological charge (assuming potentials leading to compact solitons). Next, ADM mass as well as the radius squared get negative gravitational corrections which are quadratic in the topological charge. This allowed us for a complete classification of the ADM mass-radius curves in the presence of magnetised flux. Interestingly a non-zero value of the coupling constant $\beta$ (and therefore a non-zero value of the magnetic flux) modifies {\it entirely} the constants in the parametric mass-radius formula leaving the functional form unchanged. Again as in the non-gauge case ($\beta=0$) the family of the curves can be divided into three rather distinguish groups depending where the constant $\Omega(\beta)$ is smaller, equal or bigger than 2. Another feature which is not influenced by the inclusion of the magnetic field is the fact that the maximal ADM mass is half of the maximal proper mass. 

Since the existence of the gravitating magnetised solitons is intimately related with the non gravitational case, we can conclude that there are no such solitons for double vacuum potentials (for example the so-called new baby potential). Indeed, the gravitational interaction does not have any impact on the superpotential equation. 

As all quantities rely on the knowledge on the superpotential we developed a method of a derivation of it in an approximated but analytical way. We tested such an expansion in the old baby potential case and found a perfect agreement. We believe that this approach can find some application for study of the issue of the existence of the superpotential for an arbitrary field theoretical potential $\mathcal{U}$. 

\vspace*{0.2cm}

From the physical point of view our findings tell us that the modification of the mass-radius curve can be understand as flowing the baby skyrmions towards more and more stiff matter. 

\vspace*{0.2cm}

There are many directions in which the current work can be continued. One can for example ask what happens if the Dirichlet (quadratic i.e., $\sigma$-model) term is included. Especially it would be nice to understand how this influences the mass-radius curve. Of course, due to the lack of the axial symmetry (at least for the old baby potential \cite{multi}) this can be performed only within the mean-field approximation. 

Obviously, the most important would be to investigate self-gravitating magnetized BPS  Skyrmions in 3+1 dimensions. 
\section*{Acknowledgements}
The Author thanks Andrzej Wereszczynski for discussion and Christoph Adam for comments.

\appendix
\section{Derivation of Bogomolny equations for the gauged BPS baby Skyrme model}

We start from the energy functional in the $z$ variable (\ref{Mass})
\begin{equation*}
 M=2\pi \int dz  \left(  \frac{n^2}{2e^2}a_z^2+ \frac{\lambda^2 n^2}{4} (1+a)^2 h_z^2 +\mu^2 \mathcal{U}  \right) 
\end{equation*}
We consider only static configurations. Thus,
\begin{equation}
M=2\pi  \int dz \left[ -\mathcal{L}(a,h,a_z,h_z,z)\right]
\end{equation}
where $\mathcal{L}(a,h,a_z,h_z,z)$ is Lagrangian density. We know that the Euler-Lagrange equation is invariant under the addition of total derivative $D_z F$ to $\mathcal{L}$ , where we assume that $F$ is general function of $a$ and $h$ fields. Due to that we can write down new energy density $\epsilon$ as
\begin{equation}
\label{Edensity}
\epsilon= \frac{\pi n^2}{e^2}a_z^2+ \frac{\pi \lambda^2 n^2}{2} (1+a)^2 h_z^2 +2\pi \mu^2 \mathcal{U}  - F_a a_z-F_h h_z 
\end{equation}
Now we apply FOEL method \cite{FOEL} to formula (\ref{Edensity}) and obtain the following set of FOEL equations 
\bea
&a_z&=\cfrac{e^2}{2 \pi n^2}F_a \label{1FOEL}\\
&h_z&=\cfrac{F_h}{\pi \lambda^2 n^2 (1+a)^2} \label{2FOEL} \\
&\cfrac{e^2}{4 \pi n^2}&F_a^2+\cfrac{1}{2 \pi \lambda^2 n^2 (1+a)^2}F_h^2=2\pi \mu^2 \mathcal{U} \label{SPotentialFOEL}
\eea
In order to simplify above expressions we use the fact that right side of the equation (\ref{SPotentialFOEL}) is a function of field $h$ only, which means that each component of the sum is independent of $a$
\bea
\cfrac{e^2}{4 \pi n^2}F_a^2&=&\textrm{const}(a) \\
\cfrac{1}{2 \pi \lambda^2 n^2 (1+a)^2}F_h^2&=&\textrm{const}(a)
\eea
The function $F$ that satisfies both conditions has form
\begin{equation}
\label{GDef}
F(a,h)=G(h)(1+a)
\end{equation}
where $G(h)$ is function that, after substituting (\ref{GDef}) into (\ref{1FOEL}) and (\ref{2FOEL}), obeys 
\bea
na_z&=&\cfrac{e^2}{2 \pi n}G \label{G}\\
(1+a)h_z&=&\cfrac{G_h}{\pi \lambda^2 n^2} \label{Gh} 
\eea
If we introduce new function $W(h)$ defined as
\begin{equation}
W=-\cfrac{1}{2\pi n\lambda^2}G
\end{equation}
then equations (\ref{G}) and (\ref{Gh}) take form (\ref{BOG1}) and (\ref{BOG2}).
\section{ Derivation of equation (\ref{BPSbound})}
At the beginning we start from left side of the equation (\ref{BPSbound}) in explicit form for which we use standard trick of completing a square
\begin{equation}
\int_0^{z} \tilde{\rho}(z') dz' =\int_0^{z} dz'  \left(  \frac{n^2}{2e^2}a_{z'}^2+ \frac{\lambda^2 n^2}{4} (1+a)^2 h_{z'}^2 +\mu^2 \mathcal{U}  \right)=
\end{equation}
\begin{equation}
=\int_0^{z} dz'  \left[ \cfrac{1}{2e^2}(na_{z'}+e^2\lambda^2 W)^2+\lambda^2 \left( \cfrac{n}{2}(1+a)h_{z'}+W_h \right) ^2-n\lambda^2a_{z'} W-n\lambda^2(1+a)h_{z'}W_h  \right]
\end{equation}
where we used (\ref{super}) to express potential $\mathcal{U}$ by the superpotential. Above integral can by simplified, if we use BPS equations (\ref{BOG1}) and (\ref{BOG2})
\begin{equation}
\int_0^{z} \tilde{\rho}(z') dz'=-n\lambda^2 \int_0^{z} dz'  \left[ a_{z'} W+(1+a)h_{z'}W_h  \right]=-n\lambda^2 \int_0^{z} dz' \cfrac{d}{dz'}\left[ (1+a)W\right]
\end{equation}
which, after using boundary conditions $h(0)=1$ and $a(0)=0$, give us
\begin{equation}
\int_0^{z} \tilde{\rho}(z') dz'=-n\lambda^2 \left[(1+a)W\right]|_0^z= n\lambda^2 \left[ W(1)-W(h)-a(z)W(h) \right]
\end{equation}
If $z=z_0$, then 
\begin{equation}
M=2\pi \int_0^{z_0} \tilde{\rho}(z') dz'= 2\pi n\lambda^2 W(1)
\end{equation}
which is confirmed by (\ref{Mass2}) equality.

\end{document}